# A bunching parameter interferometer: Generalization of HOM effect

Avi Marchewka

8 Galei Tchelet St., Herzliya, Israel

avi.marchewka@gmail.com

## Abstract

Are photons either bunched or unbunched, or are these particular cases of a wider phenomenon? Here we will show that bunched and unbunched photons are indeed two extreme cases of a process parameterized by a continuous parameter, called the bunching parameter, and (mainly) we will suggest a bunching interferometer that can be used for the construction and measurement of the full range of values of the above bunching parameter. Finally, as an application of the bunching parameter, we will show how the dip graph of the HOM effect is generalized.

## 1. Introduction

The exchange degeneracy symmety of identical particles gives rise to a new kind of interference, the interference between the particles' wave functions. This interference plays a role in several important quantum physics effects, e.g. the electron configuration of atoms, behavior of light, Fermi-Dirac and Boss- Einstein statistics, and many more. Among those is the bosons bunching of indistinguishing bosons (also named bosons enhancements). Bunching refers to the preference of indistinguishing bosons to be found in the same state compared to distinguishing particles under the same scenario. The footprint of bosons bunching is found in a variety of cases. (To mention a few: Brown effect [1], HOM [2], G'hosh Mandel [3], atomic optics [4]).

Feynman [5] gave a quantified measure of the bosons bunching. He showed that the probability of finding $N$ indistinguishing bosons in the same state is $N!$ higher than for $N$ distinguishing bosons (see [6])

However, it has been shown that this picture is more subtle, and in fact, Feynman's claim does not hold in general. For example, in [7] it is shown that the measure of a spatial probability of indistinguishing bosons is equal to those of distinguishing bosons. That is, the N! doesn't hold, and in fact, it is not well defined in the limiting case where the detector size goes to zero [8].

It is very tempting, as is often done, to describe the bunching of indistinguishing bosons due to "attractive forces" between the indistinguishing bosons[9]. However, this view is also only a partial truth. It has been shown [10-12] that when two bosons are released from a trap, the bosons behave as if they have "repelling forces" which govern their behavior.

Finally, a way to generalize the bosons bunching for Schrödinger particles has been given at [12]. This generalization defines a "bunching parameter", which is equal to N! in the special case considered by Feynman.

The aim of this letter is twofold. The first one, in section 2, is to formulate the bosons parameter for two photons' fields. To do this, the bunching parameter will be reformulated in the second quantization language. Then, in section 3, an interferometer will be represented with different realizations of the photons bunching parameter. This interferometer enables "tailor-made" states of arbitrary bunching parameter of photons, and, in particular, a state that is not produced in natural light. Finally, in section 4, we use those "tailor-made" states in the HOM experiment. Then, we show that such states generalize the HOM effect.

The notation of the "first quantization" follows [14] and in the "second quantization" we follow [15].

## 2. Bunching parameter for two photons.

The HOM [2] effect demonstrates clearly the bunching of two photons. In Fig 1(a), the schema of the HOM experiment is represented: two photons enter simultaneously from different legs onto a symmetric beam splitter. The notation follows [15]. For example, $|1\rangle_2$, means one particle in leg 2. The photons' probability to be found on the outcoming legs, is given at fig 1 (b) for indistinguishing photons and in fig.1(c) for distinguishing (say by their polarization degree of freedom) photons. As seen in fig1(b), the indistinguishing photons are always emitted together, whereas, as seen by fig1(c), distinguishing photons are emitted together only half of the time, and half of the time emitted to different legs. This preference of the indistinguishing bosons to emit together is a manifestation of the bosons bunching. In Fig 2 two photons enter simultaneously on the same leg of the beam splitter. In fig 2 (b) the probability of finding the emitted photons is given. It turns out that the probability of the emitted photons is independent of photons being distinguishing or not: the difference between the indistinguishing and distinguishing photons disappears. From these examples we can see that the distinguishability of the photons is not the only condition that plays a roll whether to be bunched or not.

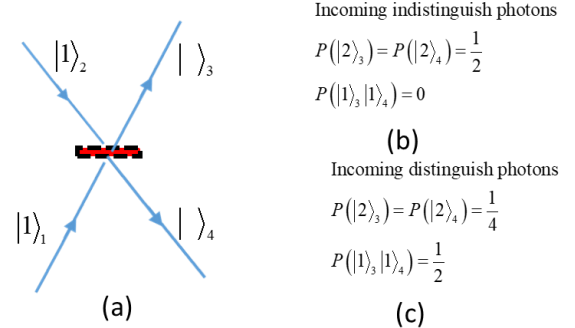

FIGURE 1: SCHEMA OF THE HOM EXPERIMENT

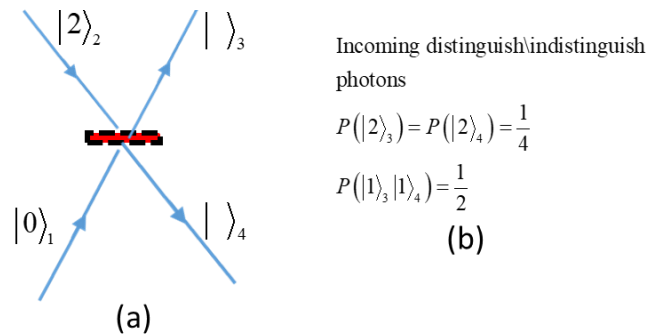

FIGURE 2: TWO PHOTONS ENTER SIMULTANEOUSLY ON THE SAME LEG

## 2.1 The bunching parameter fist quantization

Consider two particles in a two-dimensional space with an orthonormal base of two states $|q_1\rangle, |q_2\rangle$

$$|1;\chi\rangle = \sum_{i=1}^{2} \alpha_i |1;q_i\rangle$$
$$|2;\rho\rangle = \sum_{i=1}^{2} \gamma_i |2;q_i\rangle \tag{1.1}$$

With

$$\sum_{i=1}^{2} |\alpha_i|^2 = \sum_{i=1}^{2} |\gamma_i|^2 = 1 \tag{1.2}$$

The scalar product of the two states (1.1)

$$I = \langle \chi | \rho \rangle = \sum_{i=1}^{2} \alpha_i \gamma_i^* \tag{1.3}$$

Here we follow the notation of [15]. The index inside the ket $|1;\chi\rangle$ represents the particle, and the Greek later is the state the particle is in.

If the two particles are distinguishing bosons, one of the bosons is in the state $|1;\chi\rangle$ and the other is in the state $|2;\rho\rangle$, their joined wave function is,

$$|\psi\rangle^D = \frac{1}{\sqrt{N_D}} |1;\chi:2;\rho\rangle \tag{1.4}$$

Where $|1;\chi:2;\rho\rangle \equiv |1;\chi\rangle \otimes |2;\rho\rangle$ and $N_D$ is the normalization constant given by the condition $^D\langle \psi | \psi \rangle^D = 1$.

From (1.2)

$$N_D = \langle 1;\chi:2;\rho | 1;\chi:2;\rho \rangle = 1 \tag{1.5}$$

From (1.1)(1.2) (1.4) the probability for the two distinguishing bosons to be in the same state, say $|1,q_1;2,q_1\rangle$, is,

$$P^D(q_1;q_1) = |\langle 1,q_1 | 1,\chi \rangle|^2 |\langle 2,q_1 | 2,\rho \rangle|^2 = |\alpha_1 \gamma_1|^2 \tag{1.6}$$

However the joined wave function of two indistinguishing bosons has to have symmetries [14]. That is,

$$|\psi\rangle^B = \frac{1}{\sqrt{N_B}} \hat{S} |1, \chi; 2, \rho\rangle \tag{1.7}$$

Where $\hat{S}$ is the symmetric operator defined for two particles as

$$\hat{S} = \frac{1}{2}(1 + \hat{p}_{2,1}) \tag{1.8}$$

With $\hat{S}^\dagger \hat{S} = \hat{S}^2 = 1$ and $\hat{p}_{2,1}$ is the permutation operator.

Normalization of the joined bosonic wave function ${}^B\langle\psi|\psi\rangle^B = 1$ gives (1.3)

$$N_B = \langle 1, \chi; 2, \rho | \hat{S}^\dagger \hat{S} | 1, \chi; 2, \rho\rangle = \frac{1}{2}\left(1 + \langle 2, \chi; 1, \rho | \hat{p}_{2,1} | 2, \chi; 1, \rho\rangle\right) = \frac{1}{2}\left(1 + |I|^2\right) \tag{1.9}$$

That is (1.7) becomes

$$|\psi\rangle^B = \frac{(1 + \hat{p}_{2,1})}{\sqrt{2(1+|I|^2)}} |1; \chi; 2; \rho\rangle = \frac{1}{\sqrt{2(1+|I|^2)}}\left(|1; \chi; 2; \rho\rangle + |1; \rho: 2; \chi\rangle\right) \tag{1.10}$$

The probability of finding the two indistinguishing bosons in the same state, say $|1, q_1; 2, q_2\rangle$, is

$$P^B(1; q_1 : 2; q_1) = \left|\langle q_1, q_1 | \psi\rangle^B\right|^2 = \frac{2|\alpha_1 \gamma_1|^2}{(1+|I|^2)} \tag{1.11}$$

Using (1.6) and (1.11) the bunching parameter is defined by the ration

$$\beta = \frac{P^B(|q_1, q_2\rangle)}{P^D(|q_1, q_2\rangle)} = \frac{2}{1+|I|^2} \tag{1.12}$$

Before discussing the bunching parameter, we derive it in the formalism of the second quantization.

### 2.1 Bunching parameter for photons: second quantization

In the second quantization the initial state (1.1) (1.2) for distinguishing photons become,

$$|\chi\rangle = \sum_{i=1}^{2} \alpha_i \hat{a}_i^\dagger |0\rangle$$
$$|\rho\rangle = \sum_{i=1}^{2} \gamma_i \hat{b}_i^\dagger |0\rangle \tag{1.13}$$

Where one of the first photon denoted by operator $\hat{a}^\dagger$ and the second photon denoted by the operator $\hat{b}^\dagger$, and the normalization is given by (1.2).

With the following bosonic commutation relation

$$[\hat{a}_j, \hat{a}_k^\dagger] = \delta_{i,k}$$
$$[\hat{b}_j, \hat{b}_k^\dagger] = \delta_{i,k} \qquad (1.14)$$
$$[\hat{b}_j, \hat{b}_k] = [\hat{b}_j^\dagger, \hat{b}_k^\dagger] = [\hat{a}_j, \hat{a}_k] = [\hat{a}_j^\dagger, \hat{a}_k^\dagger] = 0$$

It is convenient to define

$$\hat{a}_\chi^\dagger = \sum_{i=1}^{2} \alpha_i \hat{a}_i^\dagger$$
$$\hat{b}_\rho^\dagger = \sum_{i=1}^{2} \gamma_i \hat{b}_i^\dagger \qquad (1.15)$$

The following commutation relation follows

$$[\hat{b}_\rho, \hat{a}_\chi^\dagger] = 0$$
$$[\hat{a}_\chi, \hat{a}_\chi^\dagger] = [\hat{b}_\rho, \hat{b}_\rho^\dagger] = 1 \qquad (1.16)$$

The number like operators of the states (1.15) are $\hat{N}_\chi = \hat{a}_\chi^\dagger \hat{a}_\chi$ with $\hat{N}_\chi |n\rangle = n|n\rangle$, and $\hat{N}_\rho = \hat{b}_\rho^\dagger \hat{b}_\rho$ with $\hat{N}_\rho |n\rangle = n|n\rangle$.

The joined wave function of the two distinguishing bosons is

$$|\psi\rangle^D = \frac{1}{\sqrt{N_D}} \hat{a}_\chi^\dagger \hat{b}_\rho^\dagger |0\rangle \qquad (1.17)$$

By the normalization $^D\langle\psi|\psi\rangle^D = 1$ we have .

$$N_D = \langle 0|\hat{b}_\rho \hat{a}_\chi \hat{a}_\chi^\dagger \hat{b}_\rho^\dagger|0\rangle = \langle 0|(1+\hat{N}_\chi)(1+\hat{N}_\rho)|0\rangle = 1 \qquad (1.18)$$

The probability of finding both particle in the same state, say $|1; q_1; 2; q_1\rangle$ is

$$|\langle q_1; q_1|\psi^D\rangle|^2 = |\langle 0|\hat{a}_1 \hat{b}_1 \hat{a}_\chi^\dagger \hat{b}_\rho^\dagger|0\rangle|^2 = |\alpha_1 \gamma_1|^2 \qquad (1.19)$$

If instead of the two distinguishing bosons the bosons are indistinguishing the wave function became

$$|\chi\rangle = \sum_{i=1}^{2} \alpha_i \hat{a}_i^\dagger |0\rangle$$
$$|\rho\rangle = \sum_{i=1}^{2} b_i \hat{a}_i^\dagger |0\rangle \tag{1.20}$$

Whith the bosonic commutation relation

$$\left[\hat{a}_j, \hat{a}_k^\dagger\right] = \delta_{i,k}$$
$$\left[\hat{a}_j, \hat{a}_k\right] = \left[\hat{a}_j^\dagger, \hat{a}_k^\dagger\right] = 0 \tag{1.21}$$

Accordingly we use the definition

$$\hat{a}_\chi^\dagger = \sum_{i=1}^{2} \alpha_i \hat{a}_i^\dagger$$
$$\hat{a}_\rho^\dagger = \sum_{i=1}^{2} \gamma_i \hat{a}_i^\dagger \tag{1.22}$$

The following commutation relation follows

$$\left[\hat{a}_\rho, \hat{a}_\chi^\dagger\right] = \langle \rho | \chi \rangle$$
$$\left[\hat{a}_\chi, \hat{a}_\chi^\dagger\right] = \left[\hat{a}_\rho, \hat{a}_\rho^\dagger\right] = 1 \tag{1.23}$$

The number like operator for the particles generated by (1.22) are $\hat{N}_{\chi/\rho} = \hat{a}_{\chi/\rho}^\dagger \hat{a}_{\chi/\rho}$ with $\hat{N}_{\chi/\rho}|n\rangle = n|n\rangle$.

In those terms the joined indistinguishing wave function is given by

$$|\psi\rangle^B = \frac{1}{\sqrt{N_B}} \hat{a}_\chi^\dagger \hat{a}_\rho^\dagger |0\rangle \tag{1.24}$$

Where $N_B$ is the normalization of the joined indistinguishing bosons.

Imposing the normalization ${}^B\langle \psi | \psi \rangle^B = 1$ we find

$$N_B = \langle 0 | \hat{a}_\rho \hat{a}_\chi \hat{a}_\chi^\dagger \hat{a}_\rho^\dagger | 0 \rangle = \left(1 + |\langle \chi | \rho \rangle|^2\right) = 1 + |I|^2 \tag{1.25}$$

The probability to find both indistinguishing bosons to be in the same state, say $|q_1;q_1\rangle$ with normalization $\langle q_1;q_2 | q_1;q_2 \rangle = 2$, is

$$P^B(q_1;q_2) = \left|\frac{\langle q_1;q_2 | \psi^B \rangle}{\sqrt{2}}\right|^2 = \frac{2|\alpha_1 \gamma_1|^2}{1 + |I|^2} \tag{1.26}$$

Using (1.19) and (1.26) the bunching parameter is

$$\beta = \frac{P^B(1;q_1;2;q_2)}{P^D(1;q_1;2;q_2)} = \frac{2}{1+|I|^2} \quad (1.27)$$

Equations (1.27) and (1.12) are clearly the same.

Since, $0 \leq |I|^2 \leq 1$ it follows that the bunching parameter is $1 \leq \beta \leq 2$.

It is instructive to compare this with the examples described at fig (1) and fig (2). In fig (1) the two photons have orthogonal wave function, that is $|I|^2 = 0$. It follow from equation (1.27) that $\beta = 2$ and thus

$$P^B(1;q_1;2;q_2) = 2P^D(1;q_1;2;q_2) \quad (1.28)$$

That is the probability to find the two indistinguishing bosons is twice as much as if the two bosons were indistinguishing, indeed as can be seen in fig 1(b) and Fig 1 (c).

However if the two bosons entering in the same leg, as in fig(2), then $|I|^2 = 1$. Then equation (1.27) gives $\beta = 1$. Thus,

$$P^B(1;q_1;2;q_2) = P^D(1;q_1;2;q_2) \quad (1.29)$$

That is, the probability to find the two distinguishing bosons is the same as two indistinguishing bosons, indeed as can be seen in fig 2(b).

As usual, the quantity that is invariant under unitary plays an important role. Let us show that the bunching parameter is indeed invariant under unitary transformation.

Consider two different two-dimensional spaces, with bases $|q_1;q_2\rangle$ and $|q_{1'};q_{2'}\rangle$.

The bunching parameter for the base $|q_1;q_2\rangle$ is

$$\beta = \frac{2}{1+|I|^2} \quad (1.30)$$

Likewise the bunching parameter for the base $|q_{1'};q_{2'}\rangle$ is

$$\beta' = \frac{2}{1+|I'|^2} \quad (1.31)$$

These bases are related by a unitary transformation

$$|q_{i'}\rangle = \hat{U}|q_i\rangle \quad , \quad i=1,2 \quad (1.32)$$

under which the scalar product is invariant, so $|I|^2 = |I'|^2$. Thus by (1.30) (1.31) we have $\beta' = \beta$, that is the bunching parameter is invariant under a unitary transformation.

For typical cases of emitting photons from separate sources, e.g. atoms, the photons are in orthogonal states, $|I|^2 = 0$. Since the bunching parameter is invariant under unitary transformations, it follows that to change the bunching parameter one needs an a non-unitary transformation. This will be discussed next.

3. The bunching parameter interferometry.

Due to the separate nature of atoms, two indistinguishing photons emitted by the atoms are orthogonal, $|I| = 0$. Then, their bunching parameter is $\beta = 2$. Indeed, since the original HOM [2] experiment, the bosons bunching with $\beta = 2$ has been demonstrated in many variations, e.g. [4]. This gives rise to the question of how to realize other values of the bunching parameter, i.e. $1 \leq \beta < 2$. The interferometer described in Fig(3) can be used to tail photons to have a bunching parameter with $1 \leq \beta < 2$. In Fig 3. Two incoming photons, one at the incoming legs of beam splitter $A$, and one on the incoming legs of beam splitter. $B$. Setting the delays at $| \ \rangle_{a_1}$ and at $| \ \rangle_{a_2}$ such that the photons that come from beam splitter $A$ and $B$ reach the beam $C$ splitter and beam splitter $D$ simultaneously.

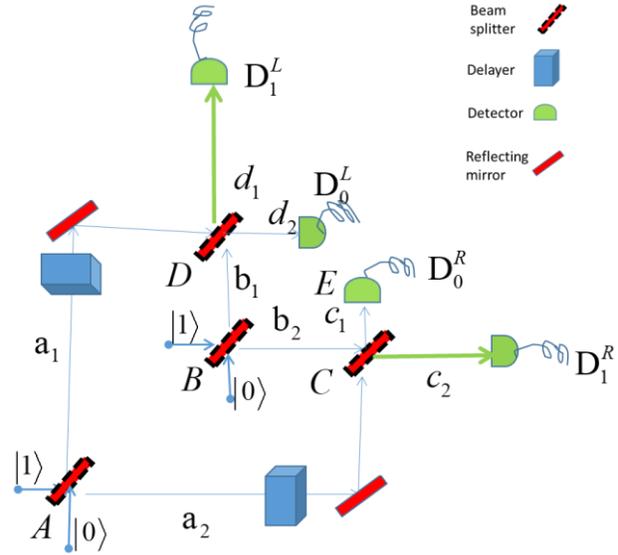

**FIGURE 3: BUNCHING PARAMETER INTERFEROMETRY**

The photons will be detected eventually in one of the four detectors $D_{0,R}, D_{0,L}, D_{1,R}, D_{1,L}$. Each of the beam splitters is unitary:

$$|t_k|^2 + |r_k|^2 = 1$$
$$t_k \bar{r}_k + \bar{t}_k r_k = 0 \tag{1.33}$$

Where $k = \{A, B, C, D\}$. The amplitude of the photons entering the beam splitter $A$ is given by

$$A: \ |1\rangle_2 |0\rangle_1 \xrightarrow{A} t_A \hat{a}^\dagger_{a_2} + r_A \hat{a}^\dagger_{a_1} \xrightarrow{C}_{D} t_C t_A \hat{a}^\dagger_{c_2} + r_C t_A \hat{a}^\dagger_{c_1} + r_D r_A \hat{a}^\dagger_{d_1} + t_D r_A \hat{a}^\dagger_{d_2} \tag{1.34}$$

Where the subscript notation is as in [15] and above. The $k$ above and below the arrow denotes the photon passes the $k$ beam splitter.

The amplitude of the photons entering the beam splitter $B$ is given by

$$B: \ |1\rangle_2|0\rangle_1 \xrightarrow{B} t_B\hat{a}^\dagger_{b_2} + r_B\hat{a}^\dagger_{b_1} \xrightarrow{C} t_C t_B \hat{a}^\dagger_{c_1} + r_C t_B \hat{a}^\dagger_{c_2} + t_D r_B \hat{a}^\dagger_{d_1} + r_D r_B \hat{a}^\dagger_{d_2} \qquad (1.35)$$

Now if the detectors $D_{0,R}, D_{0,L}$ both have zero reading, we are left with the states at legs $c_1$ and $d_1$. Such processes of condition on detectors $D_{0,R}, D_{0,L}$ are known as post selected measurements, e.g. [16].

Then the photons state at $A$ is

$$A: \ t_C t_A \hat{a}^\dagger_{c_2} + r_C t_A \hat{a}^\dagger_{c_1} + r_D r_A \hat{a}^\dagger_{d_1} + t_D r_A \hat{a}^\dagger_{d_2} \xRightarrow{D_{0,R}=D_{0,L}=0} r_C t_A \hat{a}^\dagger_{c_1} + r_D r_A \hat{a}^\dagger_{d_1} \qquad (1.36)$$

And the photons state at $B$ is

$$t_C t_B \hat{a}^\dagger_{c_2} + t_C t_B \hat{a}^\dagger_{c_1} + t_D r_B \hat{a}^\dagger_{d_1} + r_D r_B \hat{a}^\dagger_{d_2} \xrightarrow{D_{0,R}=D_{0,L}=0} t_C t_B \hat{a}^\dagger_{c_1} + t_D r_B \hat{a}^\dagger_{d_1} \qquad (1.37)$$

Accordingly, the wave functions of the photons are

$$|\psi_A\rangle = \frac{1}{\sqrt{N_1}}\left(r_C t_A \hat{a}^\dagger_{c_1} + r_D r_A \hat{a}^\dagger_{d_1}\right)|0\rangle$$

$$|\psi_B\rangle = \frac{1}{\sqrt{N_2}}\left(t_C t_B \hat{a}^\dagger_{c_1} + t_D t_B \hat{a}^\dagger_{d_1}\right)|0\rangle \qquad (1.38)$$

Where $N_1$ and $N_2$ are the normalization constants determined by the condition. $\langle\psi_A|\psi_A\rangle = \langle\psi_B|\psi_B\rangle = 1$. using the commutation relation (1.21) we find

$$N_1 = |r_C t_A|^2 + |r_D r_A|^2$$
$$N_2 = |t_C t_B|^2 + |t_D r_B|^2 \qquad (1.39)$$

Defining

$$\hat{a}^\dagger_\chi \equiv \frac{1}{\sqrt{N_1}}\left(r_C t_A \hat{a}^\dagger_{c_1} + r_D r_A \hat{a}^\dagger_{d_1}\right)$$

$$\hat{a}^\dagger_\rho \equiv \frac{1}{\sqrt{N_2}}\left(t_C t_B \hat{a}^\dagger_{c_1} + t_D r_B \hat{a}^\dagger_{d_1}\right) \qquad (1.40)$$

The joined wave function is as follows

$$|\psi\rangle^B = \frac{1}{\sqrt{N_B}} \hat{a}_\chi^\dagger \hat{a}_\rho^\dagger |0\rangle \tag{1.41}$$

we can use (1.25) to read out the overall normalization $N_B$

$$N_B = 1 + |I|^2 \tag{1.42}$$

And also, from (1.23), we have

$$I = \langle \psi_A | \psi_B \rangle = \frac{1}{\sqrt{N_1 N_2}} \left( \overline{r_C t_A} t_C t_B + \overline{r_D r_A} t_D r_B \right) \tag{1.43}$$

If however the two photons are distinguishing photons (say by their polarization) Eq. (1.34) is unchanged

$$A: \quad |1\rangle_A |0\rangle_A \xrightarrow{A} t_A \hat{a}_{a_2}^\dagger + r_A \hat{a}_{a_1}^\dagger \xrightarrow{C} t_C t_A \hat{a}_{c_2}^\dagger + r_C t_A \hat{a}_{c_1}^\dagger + r_D r_A \hat{a}_{d_1}^\dagger + t_D r_A \hat{a}_{d_2}^\dagger \tag{1.44}$$

But because the photons are distinguishing, the creation operator in (1.35) is set to $\hat{b}$

$$B: \quad |1\rangle_B |0\rangle_B \xrightarrow{B} t_B \hat{b}_{b_2}^\dagger + r_B \hat{b}_{b_1}^\dagger \xrightarrow{C} r_C t_B \hat{b}_{c_2}^\dagger + t_C t_B \hat{b}_{c_1}^\dagger + t_D r_B \hat{b}_{d_1}^\dagger + r_D r_B \hat{b}_{d_2}^\dagger \tag{1.45}$$

with the commutation relation (1.14).

The single-photon wave functions are,

$$|\psi_A\rangle^D = \frac{1}{\sqrt{N_1^D}} \left( r_C t_A \hat{a}_{c_1}^\dagger + r_D r_A \hat{a}_{d_1}^\dagger \right) |0\rangle$$

$$|\psi_B\rangle^D = \frac{1}{\sqrt{N_2^D}} \left( t_C t_B \hat{b}_{c_1}^\dagger + t_D t_B \hat{b}_{d_1}^\dagger \right) |0\rangle \tag{1.46}$$

Where $N_1$ and $N_2$ are the normalization constant determined by the condition $\langle \psi_A | \psi_A \rangle^D = \langle \psi_B | \psi_B \rangle^D = 1$. Using (1.14) gives $N_1 = N_1^D$ and $N_2 = N_2^D$.

Defining

$$\hat{a}_\chi^\dagger \equiv \frac{1}{\sqrt{N_1}} \left( r_C t_A \hat{a}_{c_1}^\dagger + r_D r_A \hat{a}_{d_1}^\dagger \right)$$

$$\hat{b}_\rho^\dagger \equiv \frac{1}{\sqrt{N_2}} \left( t_C t_B \hat{b}_{c_1}^\dagger + t_D r_B \hat{b}_{d_1}^\dagger \right) \tag{1.47}$$

Then the joined wave function of the distinguishing photons becames

$$|\psi\rangle^D = \frac{1}{\sqrt{N_D}} \hat{a}_\chi^\dagger \hat{b}_\rho^\dagger |0\rangle \tag{1.48}$$

And the normalization, $\langle\psi|\psi\rangle^2 = 1$ gives $N_D = 1$.

Using (1.19) for (1.48) and (1.26) for (1.41) the bunching parameter became

$$\beta = \frac{2}{1+|I|^2} = \frac{2}{1 + \frac{|\overline{r_C t_A} t_C t_B + \overline{r_D r_A} t_D t_B|^2}{N_1 N_2}} \qquad (1.49)$$

To see the range of values for the bunching parameter that this interferometer realizes we will consider the simplified version of that interferometer.

The general matrix for beam splitter may be represented by one parameter

$$U_{bs} = \begin{pmatrix} t' & r \\ r' & t \end{pmatrix} \qquad (1.50)$$

Such as $|r'|=|r|$, $|t'|=|t|$, $|r|^2+|t|^2=1$, $\overline{r}t'+r'\overline{t}=0$ and $\overline{r}t+r'\overline{t}'=0$.

Choosing to represent the beam splitter by a single parameter we have

$$U_{bs}(\theta_k) = \begin{pmatrix} \cos(\theta_k) & i\sin(\theta_k) \\ i\sin(\theta_k) & \cos(\theta_k) \end{pmatrix}$$
(1.51)

Where $k = \{A, B, C, D\}$ is the indexing of the beam splitter. If we chose the beam splitters $A$ and $B$ to be symmetric, $\theta_A = \theta_B = \frac{\pi}{4}$ the bunching parameter (1.49) shome in fig4. That is, for a simple setup, when the bean splitters $A$ and $B$ are symmetric, the range of the bunching parameter range is around 80% to its full range (see Fig 4)

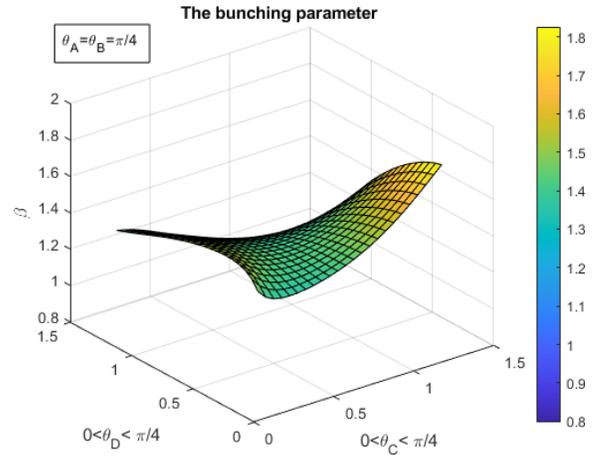

**FIGURE 4 THE BUNCHING PARAMETER RANGE**

4. Generalization of the HOME effect: an example

To see how the states orthogonality generalize the HOM effect, consider first an example. For the setup of HOM experiment Fig(2) with the following incoming photons(1.13): $a_1 = 1, a_2 = 0$ and $b_1 = 1/\sqrt{5}, b_2 = 4/\sqrt{5}$. Then, we have $|I|^2 = 1/5$ and $\beta = 1.75$. For the two incoming photons that are distinguished, the probability to find them together in one of the output legs is (symmetric beam splitter)

$$P^{(2D)} = \frac{1}{2} \quad (1.52)$$

However, the probability to find two for indistinguished photons,

$$P^{(2ID)} = \beta P^{(2D)} = \frac{5}{6} \quad (1.53)$$

It follows that the probability to find the two indistinguished photons in different legs is

$$P^{11IB} = \frac{1}{6} \quad (1.54)$$

This probability corresponds to the lowest point of the dip in HOM effect, which is not zero as in the standard HOM dip.

Let's put this in more general terms. The lowest point of the dip point corresponds to the value

$$P^{11IB} = 1 - P^{(2ID)} = 1 - \beta P^{(2D)} \quad (1.55)$$

Consider the HOM set up with $P^{(2D)} = \frac{1}{2}$ and $\beta = 2$ thus $P^{11IB} = 0$. However, in general $1 \leq \beta \leq 2$ and thus $P^{11IB} \geq 0$, which changes the lowest point of the HOM dip. More details of this will be published elsewhere.

5. Discussion and summary:

We have shown that the basic behaviors described in Fig (1) and Fig(2) [2] are specific cases of a more general behavior, parameterized by the bunching parameter. However, in natural circumstances, photons are produced from separate atoms. Then, their initial states are orthogonal, leading to the value $|I|^2 = 0$. Thus, the bunching parameter of $1 < \beta < 2$ is not an everyday phenomenon. Therefore, we introduced the bunching parameter interferometer. It has been shown that such interferometer can produce bosons with bunching factor range of 80% out of the full theoretical bunching parameter. Finally, using bosons with various bunching parameters, the generalization of the HOM dip[2] was given.

I wish to thank Dr. Oskar Pelc and Dr. Oded Kenneth for their helpful comments on the paper.